\documentclass[twocolumn,amsmath,amssymb,aps,]{revtex4-1}
\usepackage{graphicx}% Include figure files
\usepackage{dcolumn}% Align table columns on decimal point
\usepackage{bm}% bold math

\begin{document}

\title{The electron screening puzzle and nuclear clustering}% Force line breaks with \\
\author{C. Spitaleri }
\email{spitaleri@lns.infn.it}
\affiliation{Department of Physics and Astromomy, University of Catania, Catania, Italy and
INFN-Laboratori Nazionali del Sud, Catania, Italy.}

\author{C. A. Bertulani}
\email{carlos.bertulani@tamuc.edu}\affiliation{Department of Physics and Astromomy, Texas A\&M University-Commerce, Commerce, TX 75429, USA }
\affiliation{Department of Physics and Astronomy, Texas A\&M University, College Station, Texas 77843, USA}

\author{L. Fortunato}
\email{lorenzo.fortunato@pd.infn.it}
\affiliation{Dipartimento di Fisica e Astronomia ``Galileo Galilei", Universit\`a di Padova, via Marzolo, 8, I-35131 Padova, Italy}
\affiliation{ INFN, Sezione di Padova, via Marzolo, 8, I-35131 Padova, Italy }

\author{A. Vitturi}
\email{vitturi@pd.infn.it}
\affiliation{Dipartimento di Fisica e Astronomia ``Galileo Galilei", Universit\`a di Padova, via Marzolo, 8, I-35131 Padova, Italy}
\affiliation{ INFN, Sezione di Padova, via Marzolo, 8, I-35131 Padova, Italy }

\date{\today}% It is always \today, today,

\begin{abstract}
Accurate measurements of nuclear reactions of astrophysical interest within, or close to, the Gamow peak, show evidence of an unexpected effect  attributed to the presence of atomic electrons in the target. The experiments need to include an effective ``screening" potential to explain the enhancement of the cross sections at the lowest measurable energies.  Despite various theoretical studies conducted over the past 20 years and numerous experimental measurements,  a theory has not yet been found that can explain the cause of the exceedingly high values of the  screening potential needed to explain the data. In this letter we show that instead of an atomic physics solution of the ``electron screening puzzle", the reason for the large screening potential values is in fact due to clusterization effects in nuclear reactions, in particular for reaction involving light nuclei.
\end{abstract}

\pacs{Valid PACS appear here}% PACS, the Physics and Astronomy
\maketitle

To understand the energy production in stars, the first phases of the universe and the subsequent stellar evolution, an accurate knowledge of nuclear reaction cross sections  $\sigma (E)$  close to the Gamow energy E$_G$ is required \cite{Rolfs1988,Adelberger2011}. 
Therefore, recent research in experimental nuclear astrophysics has triggered the development of new theoretical methods and the introduction of new experimental techniques to study thermonuclear reactions at ultra-low energies, either directly or indirectly. 
In nuclear reactions induced by charged-particles occurring during quiescent burning in stars, E$_G$ (in general of order of few keV to 100 keV) is far below the Coulomb barrier E$^{C.B.}$ for the interacting nuclei, usually of the order of few MeV.
In particular  almost all of the nuclear reactions relevant to solar energy generation are between charged  particles and non-resonant reactions \cite{Adelberger2011}.
This implies that as energy is lowered the thermonuclear reactions are more dependent on the tunneling with an exponential decrease of the cross section.
Therefore, their bare nucleus cross sections $\sigma_b$(E) drops exponentially with decreasing energy. For such reactions it is helpful to remove the rapid energy dependence associated with the Coulomb barrier, by evaluating the probability of $s$-wave scattering off a point charge.  The nuclear physics (including effects of finite nuclear size, higher partial waves, antisymmetrization, and any atomic screening effects not otherwise explicitly treated) is then isolated in the S factor, defined by
\begin{equation}
 S(E)=E  \sigma_b(E) \exp\left[2\pi\eta(E)\right] , \label{sbe}
\end{equation}
 where $\sigma_b$(E) is the bare nucleus cross section at the center of mass energy $E$  and $\exp(2\pi\eta)$ is the inverse of the Gamow tunneling factor, which removes the dominant energy dependence of $\sigma_b(E)$ due to barrier penetrability,
and  the  Sommerfeld parameter $\eta(E)$   is defined as
\begin{equation}
\eta(E) = \frac{Z_1Z_2e^2}{\hbar v}=\frac{Z_1Z_2  \alpha}{v/c} .   \label{eta}
\end{equation}
It depends on the atomic numbers Z$_1$, Z$_2$ of the colliding nuclei and on their relative velocity $v=\sqrt{2E/\mu}$ in the entrance channel and $\alpha=e^2/\hbar c$ the fine-structure constant. Due to this definition, the astrophysical S factor is a slowly varying of $E$ and one can extrapolate $S(E)$ more reliably from the range of energies spanned by the data to the lower energies characterizing the Gamow peak. (For more details  see Refs. \cite{Adelberger2011,Barker2002} and references therein).\

The measured cross-section $\sigma_s$ must be corrected for the effect of electron screening  arising from the presence of electrons in the target atoms and, possibly, in the  (partly)-ionized projectiles \cite{Rolfs1988,Adelberger2011,Assenbaum1987,Fiorentini1995,Strieder2001}.
The presence of electrons contribute to an enhancement of the measured cross-section compared to that with bare nuclei. Note that a similar screening effect is also present in stellar plasmas, where fully ionized atoms are surrounded by a ``sea" of electrons within the so-called Debey-H\"uckel radius, which in turn depends on conditions of plasma temperature  and density that may vary during stellar evolution. Because the electron screening measured in the laboratory differs from the one in the plasma, it is important that the measured cross sections of astrophysical interest be the bare one, $\sigma_b$(E), so that plasma screening corrections can be subsequently applied.
To parameterize the cross section rise due to the screening effect, an enhancement factor $f_{lab}(E)$ is usually introduced. This factor
can be described in a simplified way by the equation \cite{Rolfs1988, Fiorentini1995, Strieder2001},
\begin{eqnarray}
f_{lab}(E) = \frac{ \sigma_s(E)}{\sigma_b(E)} = \frac{S_s(E)}{S_b(E)}  \sim \exp\left[\pi\eta\frac{U_e{^{(lab)}}}{E}\right], \label{fe}
\end{eqnarray}
where $U_e^{(lab)}$ is the electron screening potential in laboratory experiments.
A good understanding of the electron screening potential  $U_e^{(lab)}$ is essential in order to calculate $\sigma_b(E)$ from the experimental $\sigma_s(E)$ using Eq. \eqref{fe}. For astrophysical applications it is necessary to know  accurately the reaction rates in the stellar plasma, amounting to an average over the particle velocities, $\left<\sigma_{pl}(E) v\right>$. In turn, the  effective cross section  for stellar plasma,  $\sigma_{pl}$(E), 
is connected   to the bare nucleus cross section and to the stellar electron screening enhancement factor $f_{pl}$ by the relation
\begin{equation}
 \sigma_{pl}(E) = \sigma_b(E)\ f_{pl}(E) . \label{sigpl}
\end{equation}
If $\sigma_b$(E) is measured at the ultra-low energies (Gamow energy  $E_G$) and $f_{pl}(E)$ is estimated within the framework of the Debye-H\"uckel theory  it is possible from Eq. \eqref{sigpl} to evaluate  the  $\sigma_{pl}(E)$ which is the main quantity necessary  for astrophysical applications. 
Unfortunately, direct experiments to measure the cross sections  of reactions involving light nuclides have shown  that the expected enhancement of the cross section at low energies connected to the screening effect is, in many cases, significantly larger than what could be accounted for by available atomic-physics models  \cite{Rolfs1988, Strieder2001}.       
This aspect deserves special attention \cite{Barker2002} because one may have a chance to predict the effects of electron screening in an astrophysical plasma only if it is well understood under laboratory conditions (Eqs. (\ref{sbe}-\ref{fe})).
To explain the laboratory screening puzzle  many  experimental \cite{Rolfs1988,Adelberger2011,Assenbaum1987,Fiorentini1995,Strieder2001} and  theoretical studies have been carried out \cite{Barker2002, Angulo1998}.
In particular  many experiments were performed  to estimate the systematic errors in the determination of the astrophysical factor.
Special investigations were carried out to rule out errors that might be present in the extrapolation of the data to zero energy and in the calculations of the energy loss  at these ultra-low energies  \cite{Rolfs1988,Adelberger2011,Assenbaum1987,Fiorentini1995,Strieder2001}.
But up to now, theoretical studies from the point of view of atomic physics have not given a solution to this puzzle.
This lack of theoretical understanding can jeopardize the significance of some  values of the bare nucleus astrophysical factor $S_b(E)$ extracted from direct measurements.
The aim of this letter  is to uncover a  novel approach  to the solution of this puzzle by the introduction of nuclear  structure  effects  without questioning the well known atomic physics effects. The main motivation to justify the introduction of this new idea is explained next.

The wave function of a nucleus in the Fock space can be expressed as
\begin{equation}
\left |\psi_{nucleus}\right> = \alpha\left|\psi_A\right> + \beta\left|\psi_a\psi_B\right> + \gamma\left|\psi_c\psi_D\right> + \cdots ,\label{wf} 
\end{equation}
where ($\alpha, \beta, \gamma$) are spectroscopic amplitudes, $\left|\psi_A \right>$ is
the wave function of the $A$ nucleons in a non-cluster configuration and 
$\left|\psi_a\psi_B\right>$ represents the nuclear wave function in a cluster-like configuration
 with clusters a and B. Cluster configurations can alter the fusion probabilities 
because the Coulomb penetrability is reduced, as we show next. A simple evidence is the cross section for $^6$Li + $^6$Li $\to$ 3$\alpha$ reactions at ultra-low energies which are experimentally found to be orders of magnitude larger than  calculations based on barrier penetrabilities  for $^6$Li as non-clusterized spherical nuclei (\cite{Lattuada1988} and references therein).

\begin{figure}
\begin{center}
\includegraphics[scale=0.50]{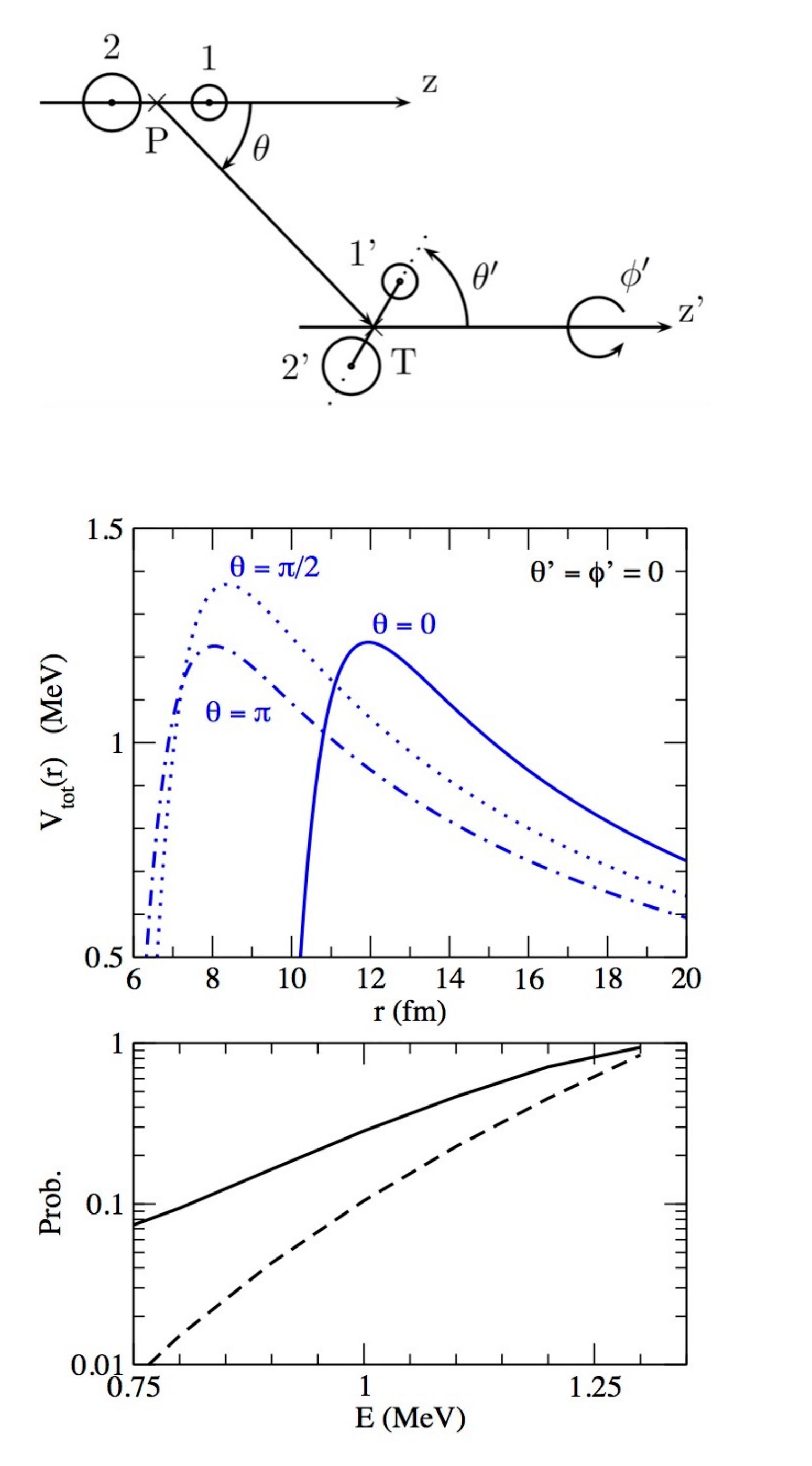}
\caption{Top figure: Coordinate system used in calculations. Plot - Upper panel: Ion-ion potentials for the dicluster systems with three different orientations as a function of c.m. distance to show the change in barrier height and position. Plot - Lower panel: Penetration probability of averaged dicluster-discluster system (solid) compared with sphere-on-sphere (dashed).\label{fign}}
\end{center}
\end{figure}

The basic idea of the cluster model and its relation to the screening puzzle is that whereas the spectroscopic amplitudes of cluster-like structures can be very small, the fusion reactions have an exponential enhancement for cluster-like structures. Since $^6$Li can have a $d + \alpha$ cluster structure, the fusion can be enhanced because the Coulomb barrier for deuterons with $^6$Li is suppressed.
Due to clustering the fusion cross section can be split into partial cross sections in the form
\begin{equation}
\sigma_L = C_{66}P_{^6{\rm Li}+^6{\rm Li}} + C_{26} P_{{\rm d}+^6{\rm Li}} + \cdots , \label{slc}
\end{equation}
where the constants $C_{66}$ and $C_{26}$ include the spectroscopic amplitudes and appropriate phase factors for $^6$Li + $^6$Li and d + $^6$Li configurations. 
Because
\begin{equation}
\frac{P_{^6{\rm Li}+^6{\rm Li}}}{P_{{\rm d}+^6{\rm Li}}} \to 0   \label{pratio}
\end{equation}
as $k \to 0$, the second term in Eq. (\ref{slc}) dominates at lower energies. Thus, even if the spectroscopic amplitudes are small, the cluster-like configuration enhances the cross section by many orders of magnitude, more than compensating the small configuration probability, also related to the preformation factor.
The experimental data clearly shows that the fusion cross section in the $^6$Li + $^6$Li $\to$ 3$\alpha$ does not decrease as fast with energy as the penetrability for $^6$Li + $^6$Li channel does  \cite{Manesse1964,Gadeken1972,Lattuada1987,Lattuada1988,Spitaleri2015}. 
An additional effect might be responsible for an even larger enhancement of the cross section. 
When the cluster-like structure in $^6$Li aligns so that the two deuterons and the two alphas are located along a line with the two deuterons closer to each other right before the reaction occurs, the barrier for the deuterons is reduced due to its larger distance to the alpha particle. Even without alignment, the average over all configurations is still reduced. We show this with a simple model based on the clusterization of $^6$Li.
Dramatic effects of clusterization  can be imprinted on the quantum tunneling probability. Consider a spherical coordinate system, depicted in Fig. \ref{fign}, placed with the origin in the center of mass of the projectile with the $z$-axis along the direction of bombardment. The $^6$Li projectile is partitioned into two clusters, a deuteron (1 in the following) and an alpha particle (indicated with 2), with centers of mass at a distance $d$ from each other on the $z$-axis, in such a way that the deuteron is on the positive side and therefore at distances $+2d/3$ and $-d/3$ respectively. The target center of mass is found at an angle $\theta$ with respect to the projectile's reference frame. The azimuthal angle of the projectile does not play any role, because it amounts to a rotation around the $z$-axis. The orientation of the generic inter-cluster axis of the target (here an identical $^6$Li system) has angles $\theta'$ and $\phi'$. 

The position, height and width of the barrier depend on the details of the potentials, sum of Coulomb and nuclear potentials, between each pair of clusters and it depends on vector distances ${\bf r}_{ij}$ as
\begin{equation}
V_{tot}(r,\theta,\theta',\phi')=\sum_{i,j=1}^2 \Biggl( \frac{Z_iZ_je^2}{r_{ij}} + V_N(r_{ij})   \Biggr),
\end{equation}
where the function parametrically depends also on the relative inter-cluster distance $d$, that we keep constant (and equal to $\sim 3.85$ fm obtained from the cluster model formula (A.4) of Ref. \cite{Mason} and data on radii \cite{Angeli}). The next level of refining of the model would include the weighting with the internal relative motion wavefunction of $^6$Li.
The upper panel of Fig. \ref{fign} shows the significant changes on the barrier height and position induced by relative rotations of the di-cluster orientation axes,  while keeping $\theta'=\phi'=0$.

For energies below the barrier, the tunneling probability can be calculated in the WKB approximation as $P=e^{-2G}$, where the Gamow tunneling factor is given by
\begin{equation}
 G(E,\theta,\theta',\phi')= \frac{\sqrt{2m}}{\hbar} \int_a^b \sqrt{V_{tot}(r,\theta,\theta',\phi') -E}~ dr \ .
\end{equation}
The angle-averaged penetration probability as a function of bombarding energy is displayed in the lower panel of Fig. \ref{fign} (solid line) and compared with the analogous calculations for sphere-on-sphere (optical potential from Ref. \cite{Pott}). 

\begin{table}
\caption{The experimental  values of the electron screening potentials, $U_e^{exp}$,
 and  theoretical adiabatic limits, $U_e$$^{adlim}$.}
\begin{center}
\begin{tabular}{lllccr}
\hline
{}&                                                                                             {}&                                               {}&                                     {}&            					          {}&                        \\[0.ex]	
{}&Reaction      				       	  	                                 {}&$U_e^{adlim}$     	             {}&$U_e^{exp}$               {}&Note    		     				  {}&Ref.                  \\[0.ex]  
{}&                  					                                     	{}& (eV)                 	             {}&(eV)                             {}&	                         		         {}&                         \\[0.ex]           
\hline
$[1]$  {}&$^2$H($d,t$)$^1$H                                                      {}&14       {}&19.1$\pm$3.4       {}&                                                  						   {}&\cite{Greife1995,Tumino2014}           \\[0.ex]       
$[2]$ {}&$^3$He($d$,$p$)$^4$He                                              {}&65        {}&109$\pm$9           {}&D$_2$ gas target          		  						   {}&\cite{Aliotta2001}    				    \\[0.ex]   
$[3]$ {}&$^3$He($d$,$p$)$^4$He                                              {}&120      {}&219$\pm$7            {}&        	         	                                         			          {}&\cite{Aliotta2001}    	 			   \\[0.ex]       
$[4]$ {}&$^3$He($^3$He,2p)$^4$He                                          {}&240     {}&305$\pm$90          {}& compilation    		                   						    {}&\cite{Adelberger2011} 			    \\[0.ex]
\hline
$[5]$  {}&$^6$Li($d$,$\alpha$)$^4$He                  		      	       {}&175     {}&330$\pm$120         {}&H gas  target      				                    				 {}&\cite{Engstler1992}    \\[0.ex]
$[6]$  {}&$^6$Li($d$,$\alpha$)$^4$He                  		              {}&175     {}&330$\pm$49           {}&	         		                                                                        {}&\cite{Engstler1992, Musumarra2001}          \\[0.ex]
$[7]$ {}&$^6$Li($p$,$\alpha$)$^3$He                 		               {}&175     {}&440$\pm$150         {}&H gas target       			 		        			        {}&\cite{Engstler1992}          \\[0.ex]
$[8]$ {}&$^6$Li($p$,$\alpha$)$^3$He                   	                       {}&175     {}&355$\pm$67         {}&        													{}&\cite{Engstler1992,Cruz,Lamia2013}           \\[0.ex]                      
$[9]$ {}&$^7$Li($p$,$\alpha$)$^4$He                     		        {}&175     {}&300$\pm$160          {}&H gas target       			 	             			               {}&\cite{Engstler1992}           \\[0.ex] 
$[10]$ {}&$^7$Li($p$,$\alpha$)$^4$He                      		        {}&175     {}&363$\pm$52          {}&      			 	                  						        {}&\cite{Engstler1992,Cruz,Lamia2012a}           \\[0.ex]   
\hline				
$[11]$  {}&$^9$Be($p$,$\alpha_0$)$^6$Li                                 {}&240       {}&788$\pm$70             		{}&  	                                        								{}&\cite{Zahnow1997,Wen2008}             \\[0.ex]                                                                                   
                                          
$[12]$  {}&$^{10}$B($p$,$\alpha_0$)$^7$                                {}&340      	{}&376$\pm$75           {}&                           										     {}& \cite{Angulo1993,Spitaleri2014}     	       \\[0.ex] 
$[13]$  {}&$^{11}$B($p$,$\alpha_0$)$^8$Be        		             {}&340     {}&447$\pm$67            	  {}&            	          	     										 {}& \cite{Angulo1993,Lamia2012}             \\[0.ex]  
\hline	
\end{tabular}
\end{center}
\end{table}

It is clear from this simple analysis that the probability is very much enhanced in  the dicluster-dicluster fusion model with respect to the sphere-on-sphere model. Thus, if the spectroscopic amplitudes in Eq. \eqref{wf} are known, the coefficients in the partial cross sections of  Eq. \eqref{slc} will also be known and the total fusion cross section will certainly display the enhancement effects due to clusterization. The problem of calculating those amplitudes is a very difficult one, not within the scope of this article. It requires a theory beyond the na\"ive shell model, which treats nuclei as a collection of nucleons. On the other hand, cluster models rely on the knowledge of preformation factors.  In this respect, ab-initio models are quite promising (see, e.g., Ref. \cite{NavQua})  but the inclusion of correlations including clusters in bound states has shown to be quite challenging and one does not seem to have reached the stage of properly assessing the values of the spectroscopic amplitudes for each cluster configuration.  The clustering effect we propose as a candidate to explain the electron screening puzzle is somewhat related to the  Oppenheimer-Phillips effect \cite{Greife1995,OpPh}, which is due to the  polarization of the deuteron in the Coulomb field of the target nucleus in deuteron induced reactions. We have shown that even without polarization  the fusion of light cluster-like nuclei can acquire enhanced tunneling when averaged over all geometric configurations. 

\begin{figure}
%\begin{center}
\includegraphics[scale=0.5]{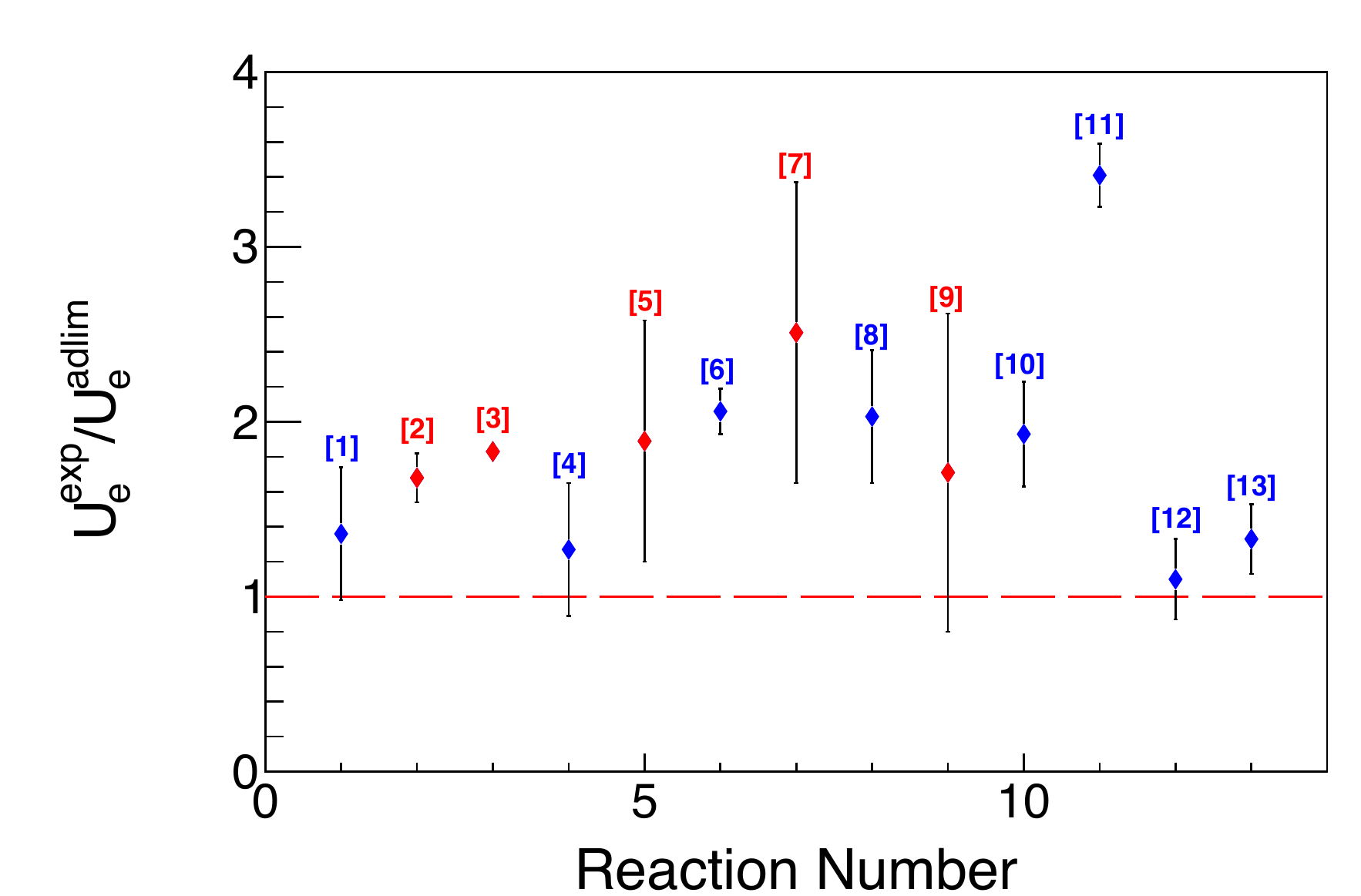}
\caption{Ratio of the experimental electron screening potential $U_e^ {exp}$  and the theoretical adiabatic limit of the electron screening potential  $U_e^ {a dlim}$  as function of the  main reaction present in the literature. 
The vertical bars are the total uncertainties of the measurements reported in literature. The numbers in brackets correspond to those in Table I.}
%\end{center}
\end{figure}

In Table I  and figure 2 we show typical cases of reactions  at ultra-low energies where clusterization fusion enhancements might be have been observed: the first is  for the case of $Z=1$ nuclei reacting with nuclei which do not present an evident nuclear cluster structure, the second is for the case of  cluster-like  nuclei.
The  main conclusion drawn from Table 1 is that there is a clear correlation between the cluster structure of nuclei involved in reactions at ultra-low energies and the discrepancy between the value of the upper limit (adiabatic approximation) of the screening potential, $U_e^{adlim}$, and  its experimental value, U$_e$$^{exp.}$. 
The disagreement increases as  the cluster structure is more pronounced (larger cluster spectroscopic factor).
In particular,  Table 1  displays the following evidences  in favor of a nuclear structure solution for the ``electron screening" puzzle for the thermonuclear reactions of astrophysical interest:
\begin{enumerate}
\item[(i) -] In all the  cases of reactions with cluster-like nuclei with one electron ($Z=1$) the experimental electron screening potentials are in agreement, within the experimental errors,  with the upper theoretical limit due to atomic energy balance \cite{Assenbaum1987} (examples are the cases of  d + d and d + p reactions).  
\item[(ii) -]  For  reactions with cluster-like nuclei  with small electron number ($Z=3$) and mass numbers 6,7 (examples are  p + $^{6,7}$Li)  if we consider the central values of  the experimental screening potential we observe that  these values are more than a factor 1.5 times higher with respect  the U$_e$$^{adlim}$. 
\item[(iii) -] Reactions with cluster nuclei with  electron number $Z=4-5$ also show a disagreement between the experimental and theoretical upper limit based on the energy balance in the adiabatic approximation. The discrepancy increases as the cluster structure of the interacting nuclei is more evident (examples are the reactions  p + $^9$Be and p + $^{10,11}$B).
\end{enumerate}
If the solution of the ``electronic screening" puzzle would be related to the effects of atomic nature the item $(ii)$ of the list above should have the same value of experimental screening potential for all three reactions (p + $^6$Li, d, p + $^6$Li and $^7$Li reactions) because of the isotopic invariance.
In fact, in the case of the reaction p + $^7$Li there is a deviation of the central value of experimental screening potential by about 250 eV, while for other cases there is a deviation of 165 eV (d + $^6$Li) and 180 eV (p + $^6$Li).
If we consider the case of the reactions in the group $(iii)$ we find that the differences found with the atomic screening predictions are $90$ eV for p + $^{10}$B, and $90-130$ eV for the reaction $^{11}$B + p.
For these latter cases, in order to draw more definitive conclusions, new measurements are needed with larger precision to reduce the margin of uncertainty.
In any case, these results are not  in agreement with the atomic description of the screening of thermonuclear reactions in the laboratory.\
The main idea introduced in  this work is that at the very low energies of nuclear reactions in stellar environments,  the condition set by Eq. \eqref{pratio} may occur
 due to the presence of clusters in the interacting  nuclei.
The Coulomb penetrability  is suppressed  with decreasing energy of interacting nuclei due to clustering and polarization.
Therefore, the absolute values of astrophysical factors obtained through extrapolation should be reviewed to include not only atomic physics but also nuclear physics effects to correct for the increase of the astrophysical S-factor as the energy decreases. 
Indeed, in Eq. \eqref{sbe} it is implicitly assumed that the wave function for the relative motion of the nuclei is expressed only by the first term of Eq. \eqref{wf} ($\alpha=1$). Therefore the Gamow factor $\exp (2\pi \eta$) is calculated taking into account that in Eq. \eqref{wf} no cluster structures exist, i.e. $\beta=\gamma = 0$.     
Only in the situation that clusters can be formed and some sort of polarization occurs, the Gamow factor can compensate for  the drastic suppression of the cross section with  decreasing energy in many of the reactions of astrophysical relevance.
These conclusions, to be confirmed by further more precise experiments, will lead to a critical review of the actual values 
of the electronic screening potentials. This problem also appears to exist with direct experiments at higher energies.
From the considerations expressed in the present work we can state that the discrepancy between the experimental electron screening potential values and the upper theoretical values (adiabatic limit) may be linked to nuclear structure effects and not to hitherto unknown and speculative processes in atomic physics.
We propose that new theoretical  and experimental  studies  in the field of nuclear astrophysics at very low energies should be carried out.
In particular, a more comprehensive theoretical reaction method that takes into account  polarization and alignment of cluster-like nuclei should be pursued.
New and more precise measurements to confirm this theory should be carried out concomitantly. 
The nuclear reactions  involving $^{6}$Li and $^{7}$Li, such as the  $^6$Li + $^6$Li,  $^7$Li + $^7$Li,  $^7$Li + $^6$Li , $^9$Be + $^3$He, $^9$Be + $^7$Li should be of particular interest to prove the relevance of such additional nuclear structure effects in thermonuclear reactions.

\par
This work has been partially supported by the Italian Ministry of University MIUR under the grant ``LNS-Astrofisica Nucleare (fondi premiali)" and the  U.S. NSF Grant No. 1415656, and U.S. DOE grant No. DE-FG02-08ER41533.
\bibliography{Screening}
\end{document}